\documentclass[11pt]{article}
\usepackage{epsfig}
\newcommand{\lmp}{Large Momentum Procedure}
\newcommand{\hmp}{Hard Mass Procedure}
\begin{document}
\title{\vskip-3cm{\baselineskip14pt
\centerline{\normalsize\hfill TTP98--25}
\centerline{\normalsize\hfill hep-ph/9806524}
\centerline{\normalsize\hfill June 1998}
}
\vskip1.5cm
Higher order corrections to $Z$--decay\thanks{Presented at the
    Zeuthen Workshop on Elementary Particle Theory ``Loops and Legs in
    Gauge Theories'', Rheinsberg, Germany, April 19--24, 1998.}
}
\author{{\sc Robert Harlander}\\[1em]
  {\small
    Institut f\"ur Theoretische Teilchenphysik,
    Universit\"at Karlsruhe,}\\ 
  {\small
    D-76128 Karlsruhe, Germany
  }}
\date{}
\maketitle
\begin{abstract}
  Recent developments in the field of automatic computation of Feynman
  diagrams using asymptotic expansions are reviewed. The hadronic decay
  rate of the $Z$--boson is taken as an example for their physical
  application.
\end{abstract}
%\PACS{12.38.Dg, 12.15.Lk, 12.38.-t, 12.38.Bx}
%\newpage
\section{\label{sec::intro}Introduction}
The final round of the analysis of data taken at LEP in the runs between
1990 and 1995 at energies around the $Z$--peak is going to be completed
\cite{clare}. About 16 millions of $Z$--bosons have been produced and the
resulting accuracy of, e.g., the $Z$--boson mass and its decay width is
impressive (for the latest experimental values of these quantities see
\cite{clare}). In view of this precision there have been huge efforts in
calculating higher order corrections to these observables. This talk is
supposed to be not so much a review of theoretical and experimental results
but is rather concerned with some of the available tools to perform such
calculations.

\section{\label{sec::asym}QCD corrections and asymptotic expansions}
The decay rate of the $Z$--boson into quarks constitutes an illuminating
example of how to use asymptotic expansions of Feynman diagrams to
simplify calculations. In what follows we will only be concerned with
the vertex corrections and it will always be assumed that they are
computed via the optical theorem by calculating the $Z$--boson self
energy and taking the imaginary part.  If one considers, e.g., only
QCD-corrections, it certainly is a reasonable lowest order approximation
to neglect all quark masses. Only one dimensional quantity is left, the
$Z$--boson mass, and one arrives at massless propagator diagrams for
which the so-called {\sl integration-by-parts} algorithm is available
and has been explicitely worked out \cite{CheTka81} and implemented in a
{\tt FORM} \cite{form} package called {\tt MINCER} \cite{mincer} up to
three loops. This immediately gives the answer for the $Z$--boson decay
rate into massless quarks up to ${\cal O}(\alpha_s^2)$
\cite{CheKatTka79DinSap79CelGon80}. Making additionally use of infrared
re-arrangement, it is even possible to extend the result to ${\cal
  O}(\alpha_s^3)$ \cite{GorKatLar91SurSam91,Che972}.

As a second step one may want to take effects induced by the quark
masses into acccount. Then, however, one is faced with two-scale Feynman
integrals. Although the full ${\cal O}(\alpha_s)$--result is known
\cite{KalSab55,BarRem73,JerLaeZer82} --- the vector part even for a very
long time ---, at ${\cal O}(\alpha_s^2)$ only certain subclasses of
diagrams have been computed analytically (e.g.~\cite{dubbub}).

For the remaining part one is forced to consider some approximation
procedure to obtain, e.g., an expansion in $m_q^2/M_Z^2$, with $m_q$ the
mass of the quark in the final state.  The recipe for the efficient
computation of such asymptotic expansions has been worked out in a series of
publications and shall not be repeated here (for a review see
\cite{Smirnov}). The essence is to expand the integrands of certain
subgraphs of the initial diagrams, leading to a complete factorization
of ``small'' and ``large'' quantities.  Two special cases may be
distinguished: When only masses appear as large quantities, the
technique is called the {\sl \hmp}. In contrast, the {\sl \lmp} deals
with the case of only large momenta.

Concerning the quark mass effects in the QCD-corrections to the
$Z$--boson decay rate, it is clear that here the appropriate procedure
is the second one.  There is only one large ($q^2=M_Z^2$) and one small
($m_q^2$) mass scale in this problem, so the factorization mentioned above
means that products of only single scale diagrams are produced: massless
propagator diagrams ($m=0$ and $q\neq 0$) and massive tadpoles ($m\neq
0$ and $q=0$).  We already mentioned the integration-by-parts algorithm
to compute the former ones. The underlying principle may also be applied
to the latter ones, as has been done in \cite{Bro92}, again up to three
loops. The implementation of this procedure has been performed, for
example, in a {\tt FORM} package named {\tt MATAD} \cite{matad}.

\begin{figure}
  \begin{center}
    \leavevmode
    \epsfxsize=6.cm
    \epsffile[105 430 560 700]{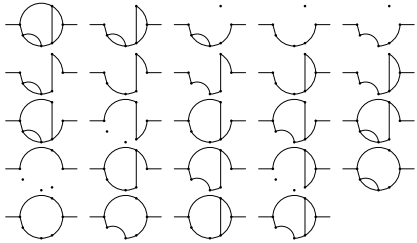}
    \hfill
%    \parbox{14.cm}{
    \caption[]{\label{O2.ps}\sloppy
      Subdiagrams contributing to the \lmp\ of a particular diagram.
      }
%}
  \end{center}
\end{figure}

There are, of course, certain limitations of this approach both from the
technical and the analytical point of view. The former one is connected
with the realization of the prescriptions provided by the asymptotic
expansions. To three-loop order it becomes a non-trivial task to find
and properly expand the contributing subdiagrams. For one particular
diagram, all subgraphs contributing to the \lmp\ are shown in
Fig.~\ref{O2.ps}.
The solution of this problem is to use the algorithmic nature of the
prescriptions and to pass their evaluation to a computer. As far as the
\lmp\ for two-point functions is concerned this has been performed in a
{\tt PERL} program called {\tt LMP} \cite{diss}.  Another program, doing
also the \hmp, will be mentioned in Section~\ref{sec::mixed}.
Therefore, in this sense this technical limitation no longer exists.

The second limitation of asymptotic expansions is more severe.  It is
clear that a cut series in general contains less information than the
full result.  In our case of the $Z$--boson decay, let us, for example,
consider the hypothetical case $2m_q<M_Z<4m_q$. The production of four
fermions is then kinematically forbidden. A small $m_q$--expansion,
however, is uncapable of discriminating among any of the cases
$M_Z>m_q$, $M_Z>2m_q$ or $M_Z>4m_q$, which is why at first sight one
cannot expect to obtain reasonable results below the four-quark
threshold. However, the situation is not so bad as it seems. The reason
is that in general the four-particle channel has a very smooth threshold
behaviour due to phase space suppression. In a small-$m_q$ expansion
this smoothness is carried over to energies below the four-particle
threshold preserving its validity in this region \cite{CheHarKueSte97}.
\begin{figure}
  \begin{center}
    \leavevmode
    \epsfxsize=6.cm
    \epsffile[110 265 465 560]{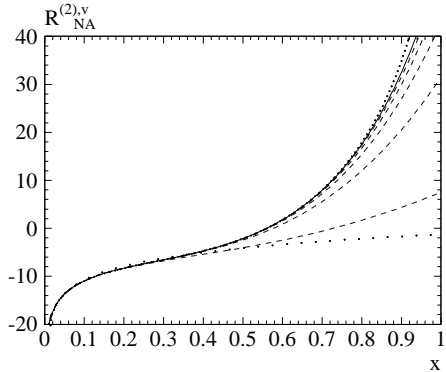} 
    \hfill
%    \parbox{14.cm}{
    \caption[]{\label{rnavx.ps}\sloppy
      Non-abelian contribution to $R_{\rm had}$ as an example for the
      application of the \lmp\ \cite{CheHarKueSte97}, plotted over
      $x=2m/\sqrt{s}$. Narrow dots: semi-analytical result from
      \cite{CheKueSte96}, other lines: power corrections up to
      $(m^2/s)^n$, $n=0,\ldots,6$.  }
%    }
\end{center}
\end{figure}

As an example the contribution to the hadronic $R$--ratio proportional to
the color factor $C_{\rm A}C_{\rm F}$ as a function of
$x=2m_q/\sqrt{s}$, where $s$ is the cms-energy, is shown in
Fig.~\ref{rnavx.ps}. The first threshold is at $\sqrt{s} = 2m_q$
($x=1$), the second one at $\sqrt{s} = 4m_q$ ($x=1/2$). Convergence,
however, seems to be warranted at least up to $x=0.8$.

Anyway, for the $Z$--boson decay one is above any possible four-quark
threshold such that convergence of the expansion to the correct answer
is guaranteed. Not only this: The quadratic and quartic mass
corrections, which already provide a very good approximation in this
case, can also be obtained without really using asymptotic expansions,
even up to ${\cal O}(\alpha_s^3)$ \cite{CheKuem2,diss}.  As far as the \lmp\ 
is concerned, there certainly are more important fields of application
for it (see, e.g., \cite{CheHarKueSte97,HarSte9712,HarSte9740}).  On the
other hand, for the \hmp, important contributions to $Z$--decay have
recently been computed where it really proves to be a very useful tool,
as we will see in Section~\ref{sec::mixed}.

\section{\label{sec::EW}Electroweak and mixed QED/QCD-corrections}
Let us turn to the electroweak radiative corrections to the $Z$--boson
decay. As far as one only considers QED, the similarity to QCD allows
not only to evaluate all pure QED-corrections, but also
QED/QCD-corrections of order $\alpha\alpha_s$,
$\alpha^2\alpha_s$ and $\alpha\alpha_s^2$ \cite{KatSur} from the knowledge
of the diagram-wise results to ${\cal O}(\alpha_s^2)$ resp.\ ${\cal
  O}(\alpha_s^3)$ by altering the color factors.

The situation is different when allowing also for $Z$-- and $W$--boson
exchange between the produced quarks, because of their non-negligible
masses. As far as the $W$--boson is concerned, another phenomenon
occurs, namely the appearence of the isospin partners of the produced
quarks as virtual particles in the loops. In contrast to the decay into
$u$--, $d$--, $s$-- or $c$--quarks, where this does not really produce a
difference in comparison to virtual $Z$--exchange because all of them
may be considered as massless, the decay into $b$--quarks is somewhat
exceptional because the $b$--quark is the isospin partner of the
$t$--quark.  Neglecting the $t$--quark mass certainly is not a good
approximation, nor can it be set infinite as one knows, for example,
from the radiative corrections to the $\rho$--parameter that appear to
be proportional to $m_t^2$ \cite{rhoparam}. Nevertheless, a full result
for $\Gamma(Z\to q\bar q)$ to ${\cal O}(\alpha)$ for the decay into
$q=u,d,s,c$ \cite{GrzKueKraStu87,BeeHol88} as well as for the one into
$b$ \cite{refzbbl1,BeeHol88} is available. The leading $m_t$--behavior
for the latter is quadratic like for the $\rho$--parameter.  To ${\cal
  O}(\alpha^2)$ the $t$--quark enters also the calculation for the decay
into $u,d,s,c$. The leading $m_t^4$-- and the subleading $m_t^2$--terms
to this order are known \cite{FleTarJeg93,DegGam97}.  For the decay into
$b$--quarks only the leading $m_t^4$--terms are available
\cite{FleTarJeg93}.

\section{\label{sec::mixed}Mixed electroweak/QCD-corrections}
A part of the mixed $\alpha\alpha_s$--corrections, namely those induced
by virtual gluon and photon exchange, has already been mentioned in
Section~\ref{sec::EW}. The present section will be concerned with the
case of $W$-- or $Z$--, accompanied by an additional gluon-exchange,
i.e., in a sense, QCD-corrections to the ${\cal O}(\alpha)$--results
described in Section~\ref{sec::EW}. Again it is natural to distinguish
between the decay into $u,d,s,c$ and into $b$. The former case was
evaluated in \cite{CzaKue96}, and it is instructive to dwell a bit on
the technique which was used in this work.

Again the rate was determined by computing the $Z$--boson self energy up
to the order considered and taking the imaginary part of the result. In
the case of virtual $Z$--boson exchange, one arrives at three-loop
on-shell integrals, for $W$--exchange one gets two-scale diagrams. The
idea in \cite{CzaKue96} was to use the \hmp\ described in
Section~\ref{sec::asym} to expand the diagrams in $M_Z^2/M^2$, where $M$
is the mass of the virtual gauge boson, and take the limit $M\to M_Z$
resp.\ $M\to M_W$ in the final result. Since convergence at these points
turned out to be quite slow, a part of the diagrams was also expanded in
$M^2/M_Z^2$. In this way it was possible to obtain a reasonable
approximation to the full result.

In the case of $Z\to b\bar b$ one faces the problem of an additional
mass scale, the $t$-quark mass. Using the \hmp\ for $m_t^2\gg
M_Z^2,M_W^2$, one may factor out the $m_t$--dependence. However, for a
part of the diagrams one still is left with two-scale and even
three-scale integrals involving $M_Z^2$ and $M_W^2$ and $\xi_W M_W^2$, where
$\xi_W$ is the electroweak gauge parameter which we want to keep.
Although they appear to be only one-loop integrals, their exact
evaluation up to ${\cal O}(\epsilon)$ produces inconvenient results.
Instead, the results of \cite{HarSeiSte97} were obtained by applying the
\hmp\ to these kinds of diagrams once more, this time using $\xi_W
M_W^2, M_W^2\gg
M_Z^2$. This seemingly unrealistic choice of scales becomes justified by
recalling the discussion of Section~\ref{sec::asym}: It is not possible
for an expansion to distinguish the inequality $M_W^2\gg M_Z^2$ from
$4M_W^2\gg M_Z^2$ or $(m_t+M_W)^2\gg M_Z^2$, the latter ones being
perfectly alright. The only matter is to perform the expansion on the
appropriate side of all thresholds, and here one is concerned with
thresholds at $2M_W$ and at $m_t+M_W$. Therefore, the choice $M_W^2\gg
M_Z^2$ is to be understood purely in this technical sense. Graphically
this continued expansion looks as follows:
\begin{eqnarray*}
&& \epsfxsize=5em 
\raisebox{-1.3em}{\epsffile[120 260 460 450]{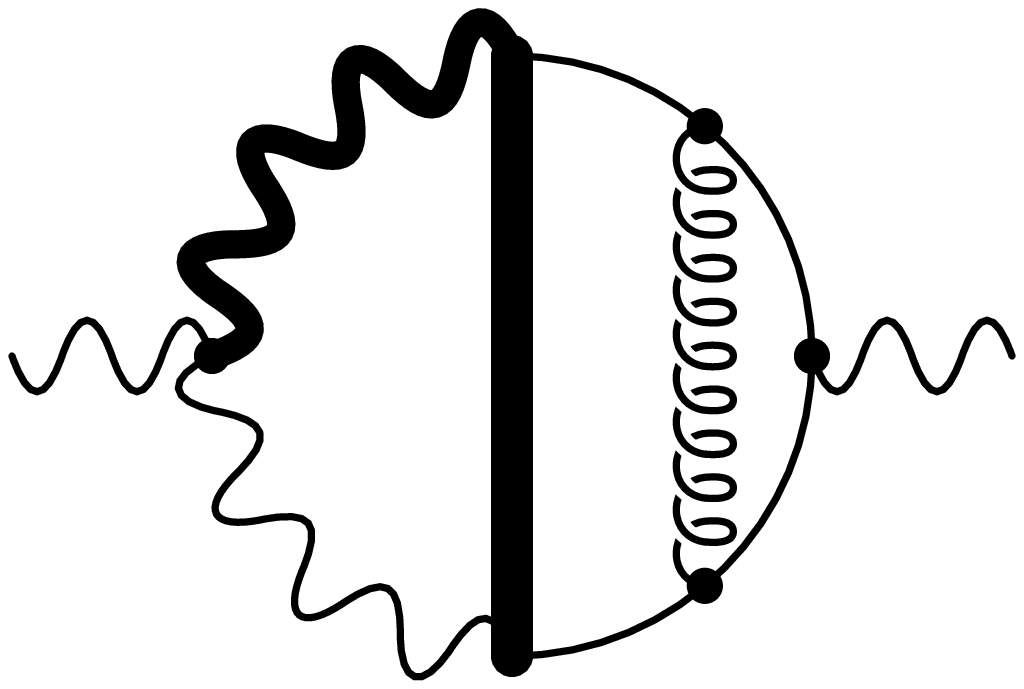}}
\stackrel{m_t^2\to \infty}{\longrightarrow}
\,\,\,\,\,
\epsfxsize=6em \raisebox{-1.3em}{\epsffile[120 260 560 450]{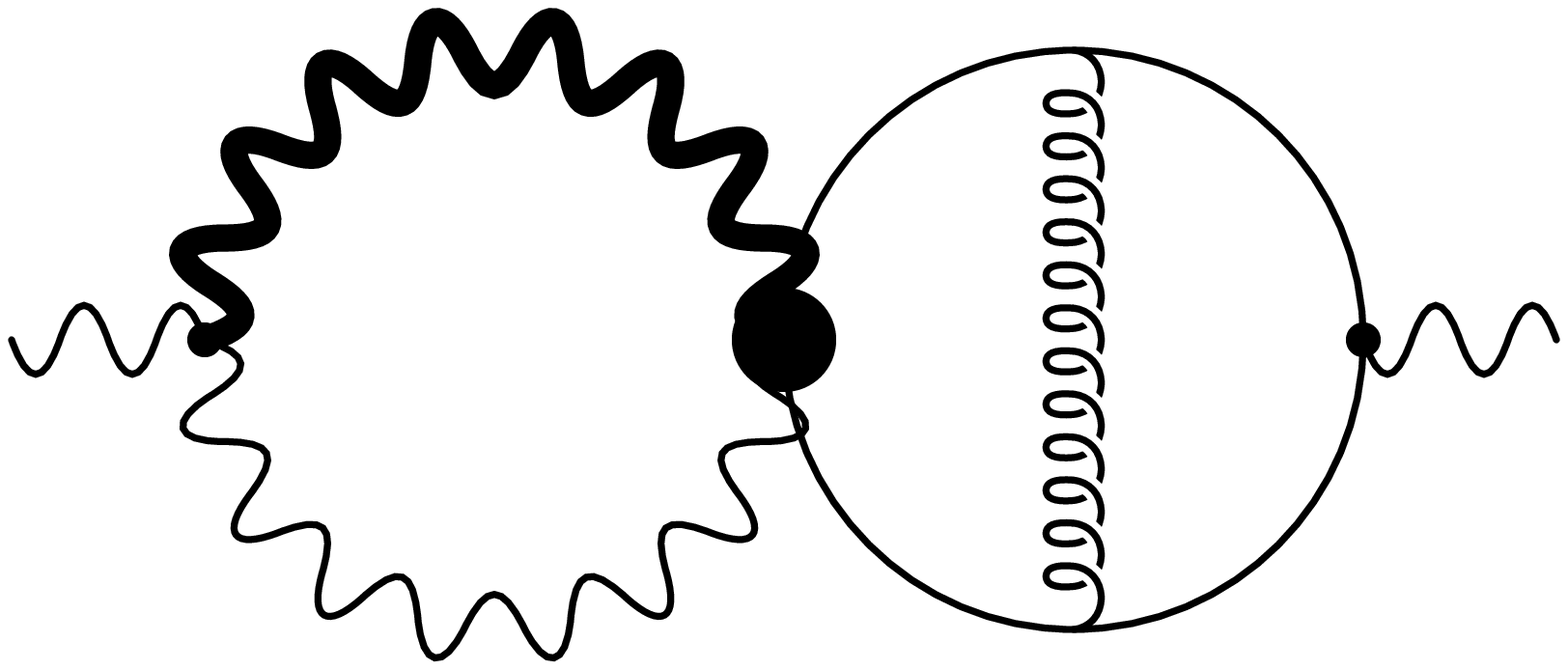}} 
  \star
\epsfxsize=.7em 
\raisebox{-1.3em}{\epsffile[260 260 310 450]{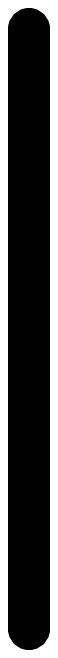}} 
\,\,+ \,\,\cdots \\[.5em]&&\hspace{6em}
\stackrel{\xi M_W^2\to \infty}{\longrightarrow}
  \bigg(\!\!\!\!
  \epsfxsize=5em
  \raisebox{-1.3em}{\epsffile[120 260 460 450]{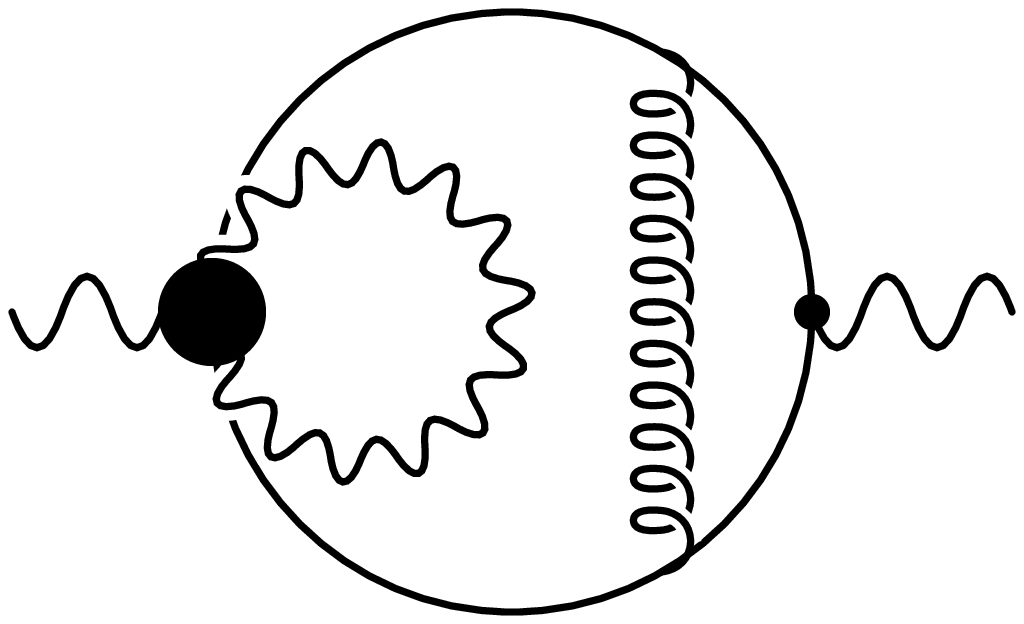}}
  \star 
  \epsfxsize=4em \raisebox{-1.em}{\epsffile[120 260 460
  450]{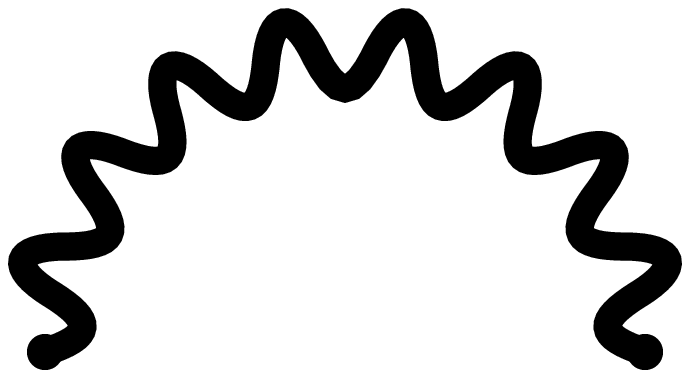}} \hspace{-2em}\bigg)
\star
\epsfxsize=.7em 
\raisebox{-1.3em}{\epsffile[260 260 310 450]{T11hmp.ps}} + \cdots\,,
\end{eqnarray*}
where only those terms are displayed which are relevant in the
discussion above and all others contributing to the \hmp\ are merged
into the ellipse.  The thick plain line is the top quark, the thick wavy one
a Goldstone boson with mass squared $\xi_W M_W^2$, for example.  The thin
plain lines are $b$--quarks, the inner thin wavy lines are $W$--bosons,
the outer ones $Z$-bosons. The spring-line is a gluon. The mass
hierarchy is assumed to be $m_t^2\gg \xi_W M_W^2 \gg M_W^2 \gg
M_Z^2$. The freedom in choosing the magnitude of $\xi_W$ provides a
welcome check of the routines and the results.

The outcome of this procedure is a nested series: The coefficients of
the $M_W/m_t$--expansion are in turn series in $M_Z/M_W$. Note that in
contrast to the decay into $u,d,s,c$ there is no threshold at $M_W$
which makes an additional expansion in $M_W/M_Z$ unnecessary.

In view of this calculation the procedure of successive application of
the \hmp\ resp.\ the \lmp\ has been implemented in a Fortran 90 program
named {\tt EXP} \cite{SeiDiplom}. Therefore, the computation of a
three-loop two-point function can now be done fully automatically given
some arbitrary hierarchy of mass scales. Even more, the link to the
Feynman diagram generator {\tt QGRAF} \cite{qgraf} in a common
environment called {\tt GEFICOM} \cite{geficom} allows to obtain the
result of a whole physical process without any human interference except
for specification of the process and final renormalization.

Finally, let us present the result for the $W$--induced corrections to
the $Z$--decay rate $\delta\Gamma^W(Z\to b\bar b)$ in the form of the
renormalization scheme independent difference to the decay rate into
$d\bar d$. Inserting the on-shell top mass $m_t = 175$~GeV,
the $Z$--mass $M_Z=91.91$~GeV and $\sin^2\theta_W = 0.223$ gives
\begin{eqnarray}
&&\delta\Gamma^W(Z\to b\bar b) - \delta\Gamma^W(Z\to d\bar d) =
    \Gamma^0 {1\over \sin^2\theta_W}
    {\alpha\over \pi}
  \bigg\{ - 0.50 
  \nonumber\\[.5em]&&\hbox{\hspace{1em}} 
  + (0.71 -0.48)+ (0.08 - 0.29) + (-0.01 - 0.07) + (-0.007 - 0.006) 
  \nonumber\\[.5em]&&\mbox{}
  + {\alpha_s\over \pi} \bigg[ 1.16 + (1.21 - 0.49) + (0.30 - 0.65) +
  (0.02 - 0.21 + 0.01) 
\nonumber\\&&\mbox{\hspace{1em}} 
+ (-0.01 - 0.04
  + 0.004) \bigg] \bigg\} = 
\nonumber\\&&\mbox{\hspace{0em}}
=\Gamma^0 {1\over \sin^2\theta_W} {\alpha\over \pi} \bigg\{- 0.50 - 0.07 +
{\alpha_s\over \pi} \bigg[ 1.16 + 0.13 \bigg]\bigg\}\,,
\label{eqgamnum}
\end{eqnarray}
where the factor $\Gamma^0\alpha/(\pi\sin^2\theta_W)$ with $\Gamma^0 =
M_Z \alpha/(4 \sin^2\theta_W \cos^2\theta_W)$ has been pulled out for
convenience. The numbers after the first equality sign correspond to
successively increasing orders in $1/m_t^2$, where the brackets collect
the corresponding constant, $\log m_t$ and, if present, $\log^2
m_t$--terms. The numbers after the second equality sign represent the
leading $m_t^2$--term and the sum of the subleading ones.  The ${\cal
  O}(\alpha)$ and ${\cal O}(\alpha\alpha_s)$--results are displayed
separately. Comparison of this expansion of the one-loop terms to the
exact result of \cite{BeeHol88} shows agreement up to $0.01\%$
which gives quite some confidence in the $\alpha\alpha_s$--contribution.
One can see that although the $m_t^2$--, $m_t^0$-- and $m_t^0\log
m_t$--terms are of the same order of magnitude, the final result is
surprisingly well represented by the $m_t^2$--term, since the subleading
terms largely cancel among each other.\\[1em]
{\bf Acknowledgements:}
I would like to thank K.G.~Chetyrkin, J.H.~K\"uhn, M.~Steinhauser and
T.~Seidensticker for fruitful collaboration, and the organizers and
participants at the workshop for the enjoyable atmosphere.
This work was supported by the ``Landesgraduiertenf\"orderung'' and the
``Graduiertenkolleg Elementarteilchenphysik'' at the University of Karlsruhe.

\def\app#1#2#3{{\it Act.\ Phys.\ Pol.\ }{\bf B #1} (#2) #3}
\def\apa#1#2#3{{\it Act.\ Phys.\ Austr.\ }{\bf#1} (#2) #3}
\def\fortp#1#2#3{{\it Fortschr.\ Phys.\ }{\bf#1} (#2) #3}
\def\npb#1#2#3{{\it Nucl.\ Phys.\ }{\bf B #1} (#2) #3}
\def\plb#1#2#3{{\it Phys.\ Lett.\ }{\bf B #1} (#2) #3}
\def\prd#1#2#3{{\it Phys.\ Rev.\ }{\bf D #1} (#2) #3}
\def\pR#1#2#3{{\it Phys.\ Rev.\ }{\bf #1} (#2) #3}
\def\prl#1#2#3{{\it Phys.\ Rev.\ Lett.\ }{\bf #1} (#2) #3}
\def\prc#1#2#3{{\it Phys.\ Reports }{\bf #1} (#2) #3}
\def\cpc#1#2#3{{\it Comp.\ Phys.\ Commun.\ }{\bf #1} (#2) #3}
\def\nima#1#2#3{{\it Nucl.\ Inst.\ Meth.\ }{\bf A #1} (#2) #3}
\def\pr#1#2#3{{\it Phys.\ Reports }{\bf #1} (#2) #3}
\def\sovnp#1#2#3{{\it Sov.\ J.\ Nucl.\ Phys.\ }{\bf #1} (#2) #3}
\def\yadfiz#1#2#3{{\it Yad.\ Fiz.\ }{\bf #1} (#2) #3}
\def\jetp#1#2#3{{\it JETP\ Lett.\ }{\bf #1} (#2) #3}
\def\zpc#1#2#3{{\it Z.\ Phys.\ }{\bf C #1} (#2) #3}
\def\epjc#1#2#3{{\it Eur.\ Phys.\ J.\ }{\bf C #1} (#2) #3}
\def\ptp#1#2#3{{\it Prog.\ Theor.\ Phys.\ }{\bf #1} (#2) #3}
\def\nca#1#2#3{{\it Nuovo\ Cim.\ }{\bf #1A} (#2) #3}
\def\mpl#1#2#3{{\it Mod.\ Phys.\ Lett.\ }{\bf A #1} (#2) #3}
\def\tmf#1#2#3{{\it Teor.\ Mat.\ Fiz.\ }{\bf #1} (#2) #3}
\def\ibid#1#2#3{{ibid.\ }{\bf #1} (#2) #3}
\def\cmp#1#2#3{{\it Comm.\ Math.\ Phys.\ }{\bf #1} (#2) #3}
\def\jcp#1#2#3{{\it J.\ Comp.\ Phys.\ }{\bf #1} (#2) #3}


\begin{thebibliography}{99}
\bibitem{clare} R. Clare, {\it LEP Physics Results}, these proceedings.
\bibitem{CheTka81}
F.V. Tkachov, \plb{100}{1981}{65};\\
K.G. Chetyrkin and F.V. Tkachov, \npb{192}{1981}{159}.
\bibitem{form}
J.A.M. Vermaseren, {\it Symbolic Manipulation with FORM}
(Computer Algebra Netherlands, Amsterdam, 1991).
\bibitem{mincer} 
S.A. Larin, F.V. Tkachov and J.A.M. Vermaseren,
Rep.~No.~NIKHEF-H/91-18 (Amsterdam, 1991).
\bibitem{CheKatTka79DinSap79CelGon80} K.G. Chetyrkin, A.L. Kataev and
  F.V. Tkachov, \plb{85}{1979}{277};\\ M. Dine and J. Sapirstein,
  \prl{43}{1979}{668};\\ W. Celmaster and R.J. Gonsalves,
  \prl{44}{1980}{560}.
\bibitem{GorKatLar91SurSam91}
S.G. Gorishny, A.L. Kataev and S.A. Larin, \plb{259}{1991}{144};\\
L.R. Surguladze and M.A. Samuel, \prl{66}{1991}{560}; (E) ibid., 2416.
\bibitem{Che972} K.G. Chetyrkin, \plb{404}{1997}{161}.
\bibitem{KalSab55} G. K\"allen and A. Sabry, {\it
    K. Dan. Videnk. Selsk. Mat.-Fys. Medd.} {\bf 29} (1955) No.~17.
  \bibitem{BarRem73} R. Barbieri and E. Remiddi, \nca{13}{1973}{99}.
\bibitem{JerLaeZer82}
J. Jers\'ak, E. Laermann and P. Zerwas,
\prd{25}{1982}{1218}; (E) \ibid{D 36}{1987}{310}.
\bibitem{dubbub}
A.H. Hoang, M. Je\.zabek, J.H. K\"uhn and T. Teubner, \plb{338}{1994}{330};\\
A.H. Hoang, J.H. K\"uhn and T. Teubner, \npb{452}{1995}{173};\\
A.H. Hoang and T. Teubner, Rep.~Nos.~DTP/97/68,
  UCSD/PHT~97-16 (Durham, San Diego, 1997), hep-ph/9707496.
\bibitem{Smirnov} V.A.~Smirnov, {\it Renormalization and Asymptotic
    Expansion} (Birkh\"auser, Basel, 1991); \mpl{10}{1995}{1485}.
\bibitem{Bro92} D.J. Broadhurst, \zpc{54}{1992}{54}.
\bibitem{matad} M. Steinhauser, Dissertation (Shaker Verlag, Aachen, 1996).
\bibitem{diss} R. Harlander, Dissertation at the University of
  Karlsruhe, 1998, to be published.
\bibitem{CheHarKueSte97}
K.G. Chetyrkin, R. Harlander, J.H. K\"uhn and M. Steinhauser,
\npb{503}{1997}{339}.
\bibitem{CheKueSte96}
  K.G. Chetyrkin, J.H. K\"uhn and M. Steinhauser, \plb{371}{1996}{93};
  \npb{482}{1996}{213}.
\bibitem{CheKuem2}
K.G. Chetyrkin and J.H. K\"uhn, \npb{432}{1994}{337};
\plb{248}{1990}{359};
\plb{406}{1997}{102}.
\bibitem{HarSte9712}
  R. Harlander and M. Steinhauser, \prd{56}{1997}{3980}.
\bibitem{HarSte9740}
  R. Harlander and M. Steinhauser, \epjc{2}{1998}{151}.
\bibitem{KatSur}
A.L. Kataev, {\it Phys. Lett.} {\bf B 287} (1992) 209;\\
L.R. Surguladze, Rep.~No.~UAHEP-969 (Alabama, 1998), hep-ph/9803211.
\bibitem{rhoparam}
  D.A. Ross and M. Veltman, \npb{95}{1975}{135};\\
  M. Veltman, \npb{123}{1977}{89}.
\bibitem{GrzKueKraStu87} B. Grzadkowski, J.H. K\"uhn, P. Krawczyk and
  R.G. Stuart, \npb{281}{1987}{18}.
\bibitem{BeeHol88} W. Beenakker and W. Hollik, \zpc{40}{1988}{141}.
\bibitem{refzbbl1} 
  A. Akhundov, D. Bardin and T. Riemann, \npb{276}{1986}{1};\\ 
  J. Bernabeu, A. Pich and A. Santamaria, \plb{200}{1988}{569};\\ 
  B.W. Lynn and R.G. Stuart, \plb{252}{1990}{676}.
\bibitem{FleTarJeg93}
R. Barbieri, M. Beccaria, P. Ciafaloni, G. Curci and A. Vicere,
\plb{288}{1992}{95}; (E) \ibid{B 312}{511};
\npb{409}{1993}{105};\\
J. Fleischer, O.V. Tarasov and F. Jegerlehner, \plb{319}{1993}{249}.
\bibitem{DegGam97} G. Degrassi, S. Fanchiotti and  A. Sirlin,
  \npb{351}{1991}{49};\\
  G. Degrassi and A. Sirlin, \npb{352}{1991}{342};\\
  G. Degrassi, P. Gambino and A. Sirlin, \plb{394}{1997}{188};\\
  G. Degrassi, P. Gambino and A. Vicini, \plb{383}{1996}{219}.
\bibitem{CzaKue96}
A. Czarnecki and  J.H. K\"uhn, {\it Phys. Rev. Lett.} {\bf 77} (1996)
3955; hep-ph/9608366(v2).
\bibitem{HarSeiSte97}
  R. Harlander, T. Seidensticker and M. Steinhauser, \plb{426}{1998}{125}.
\bibitem{SeiDiplom} T. Seidensticker, Diploma Thesis (Karlsruhe, 1998),
  unpublished.
\bibitem{qgraf} P. Nogueira, \jcp{105}{1993}{279}.
\bibitem{geficom} K.G. Chetyrkin and M. Steinhauser, unpublished.
\end{thebibliography}
\end{document}